\documentclass[conference]{IEEEtran}
\IEEEoverridecommandlockouts
\usepackage{url}
\usepackage{amsmath,amssymb,amsfonts}
\usepackage{algorithmic}
\usepackage{graphicx}
\usepackage{textcomp}
\usepackage{multirow}
\usepackage{xcolor}
\def\BibTeX{{\rm B\kern-.05em{\sc i\kern-.025em b}\kern-.08em
    T\kern-.1667em\lower.7ex\hbox{E}\kern-.125emX}}
\usepackage{tikz}
\usetikzlibrary{shapes.geometric,arrows, fit,positioning}
\usetikzlibrary{arrows.meta}

\usepackage{pgfplots}
\usepackage{pgfplotstable}
\usepackage[square, sort]{natbib}
\pgfplotsset{compat=1.18} 
\begin{document}
\pagenumbering{arabic} 
\title{Evaluating the Performance of a  D-Wave Quantum Annealing System for Feature Subset Selection in Software Defect Prediction\\
}

\author{
    \IEEEauthorblockN{Ashis Kumar Mandal\IEEEauthorrefmark{1}, 
    Md Nadim\IEEEauthorrefmark{1}, 
    Chanchal K. Roy\IEEEauthorrefmark{2}, 
    Banani Roy\IEEEauthorrefmark{2}, 
    Kevin A. Schneider\IEEEauthorrefmark{2}}
    \IEEEauthorblockA{
        Department of Computer Science, University of Saskatchewan, Saskatoon, Canada \\
        Email: \IEEEauthorrefmark{1}\{ashis.62, mnadims.cse\}@gmail.com,
        \IEEEauthorrefmark{2}\{chanchal.roy, banani.roy, kevin.schneider\}@usask.ca
    }
}

\maketitle

\begin{abstract}
Predicting software defects early in the development process not only enhances the quality and reliability of the software but also decreases the cost of development. A wide range of machine learning techniques can be employed to create software defect prediction models, but the effectiveness and accuracy of these models are often influenced by the choice of appropriate feature subset. Since finding the optimal feature subset is computationally intensive, heuristic and metaheuristic approaches are commonly employed to identify near-optimal solutions within a reasonable time frame. Recently, the quantum computing paradigm quantum annealing (QA) has been deployed to find solutions to complex optimization problems.  This opens up the possibility of addressing the feature subset selection problem with a QA machine. Although several strategies have been proposed for feature subset selection using a QA machine, little exploration has been done regarding the viability of a QA machine for feature subset selection in software defect prediction.  This study investigates the potential of D-Wave QA system for this task, where we formulate a mutual information (MI)-based filter approach as an optimization problem and utilize a D-Wave Quantum Processing Unit (QPU) solver as a QA solver for feature subset selection. We evaluate the performance of this approach using multiple software defect datasets from the AEEM, JIRA, and NASA projects. We also utilize a D-Wave classical solver for comparative analysis. Our experimental results demonstrate that QA-based feature subset selection can enhance software defect prediction. Although the D-Wave QPU solver exhibits competitive prediction performance with the classical solver in software defect prediction, it significantly reduces the time required to identify the best feature subset compared to its classical counterpart. 

\end{abstract}


\section{Introduction}
Identifying and correcting defects in software projects is one of the key aspects of Software Quality Assurance (SQA) \cite{MAXIM201619}. Defect prediction is the process of developing a model that software practitioners can use to identify defective classes or modules before the testing phase \cite{pandey2021machine}. 
It is necessary in modern software development because software defects can cause various problems, such as system crashes, data corruption, and security vulnerabilities \cite{shin2013can}. To tackle the software defect prediction task, statistical analysis, machine learning techniques, and, more recently, deep learning approaches have been used \cite{omri2020deep}. However, classification models alone are not sufficient for the task of predicting software defects. Feature subset selection algorithms also play a helpful role in finding the appropriate feature subset that can avoid irrelevant and redundant features, improve overfitting, and simplify the classification model. Since finding the appropriate feature subset is a combinatorial optimization problem, it requires a significant amount of computational time to identify the optimal feature subsets through an exhaustive search. While various search strategies, including sequential search, heuristic, and metaheuristic search, are frequently employed to identify the optimal feature subset solution in reasonable time \cite{nguyen2020survey}, these strategies tend to get stuck in a local optimum and produce sub-optimal solutions. In the context of feature subset selection for software defect prediction tasks, numerous strategies have been implemented \cite{ali2023analysis}. However, the literature suggests that all strategies used for feature subset selection in software defect prediction are executed in a traditional computing environment.
  
Quantum computing has recently emerged as a new computational paradigm offering significant computational advantages over classical computing for specific problems \cite{imre2014quantum}. One type of quantum computing is the D-Wave quantum annealing (QA) system, which specializes in solving optimization problems and related combinatorial optimization problems \cite{khairy2020learning}. Although some feature subset selection approaches in quantum computing have been recently examined \cite{ali2023analysis}, the potential of QA in tackling feature subset selection for software defect prediction remains largely unexplored.

This research study aims to assess how well a D-Wave QA system performs in selecting optimum feature subsets to predict software defects. D-Wave Quantum Processing Unit (QPU)  solver has been used as the QA solver for the task. The following research questions are posed to fulfill this aim:

\begin{itemize}
    \item  RQ1: How can a quadratic unconstrained binary optimization (QUBO) objective function be formulated to optimize feature subset selection?

    \item RQ2: To what extent does feature subset selection performed by a D-Wave QA solver  reduce feature cardinality in software defect prediction datasets?

     \item RQ3: How does the computational time needed for feature subset selection using a  D-Wave QA solver  compare with that required by a classical solver?
    
    \item RQ4: How does feature subset selection using a D-Wave QA solver compare to its classical solver regarding  prediction performance?

\end{itemize}

The main contribution of this study is as follows.

\begin{itemize}
    \item To our understanding, this is the inaugural study to explore the potential of feature subset selection in software defect prediction using D-Wave QA solver.

    \item We have examined a different array of defect prediction datasets and use a MI-based QUBO objective function to transform the datasets for the purpose of executing feature subset selection on a QA system.

    \item We have evaluated the performance of the feature subset derived from D-Wave  QA solver for software defect prediction and juxtaposed it with the performance of its classical solver. 
\end{itemize}

\section{Background}

\textbf{Quantum computing:} 
The fundamental unit of quantum computing is a qubit, representing the superposition of classical zero and one states. Quantum computers can process information in parallel by leveraging the principles of superposition, quantum entanglement, and interference \cite{deutsch1985quantum}. 
Owing to their potential to solve specific problems faster than classical computers, numerous leading IT companies have developed quantum computers based on various quantum computing models. The two primary models are QA \cite{finnila1994quantum} and the gate model \cite{hagouel2012quantum}. The gate model, also known as the universal model, can solve virtually all types of problems. Gate-based quantum computers, such as IBM’s gate model quantum computers \cite{srinivasan2018efficient}, perform computations using a sequence of quantum gates. On the other hand, the QA paradigm specializes in solving optimization problems. It seeks the lowest-energy solution (global minimum) for optimization problems encoded within the Ising or QUBO mathematical models, as exemplified by D-Wave Systems’ model \cite{grant2020adiabatic}. The ultimate goal of both paradigms is to harness immense power to address numerous complex real-world problems. 
 
\textbf{Quantum annealing (QA) Machine:} A QA machine is a quantum computing model used to find the lowest energy state of a system, which corresponds to the optimal solution of an optimization problem \cite{PhysRevApplied.5.034007}.  To solve an optimization problem using QA machine, a crucial step is to transform the problem into a QUBO problem \cite{PhysRevApplied.18.034016}. The QA process operates by encoding the QUBO formulation into the energy function (Hamiltonian) of a quantum system and then solving the problem by finding the lowest energy state.

In QUBO problem, the objective is to discover a binary vector ‘$X$’ that optimize (e.g., minimizes) the objective function. As the number of variables in a binary vector increases, the search space expands exponentially. This makes solving a QUBO problem computationally demanding in classical computing. In this scenario, the QA machine can efficiently discover the optimal binary vector. The mathematical representation of a QUBO problem in minimization is as follows:

\begin{equation}
    \text{Minimize} \quad f(X) = \sum_{i} q_{ii}x_i + \sum_{i<j} q_{ij}x_ix_j
\end{equation}

Where $X$ denotes a vector of binary variables where each $x_i$ can take a value of 0 or 1. The term $q_{ii}$ signifies the linear coefficients associated with each individual variable $x_i$. In the quadratic component of the equation the coefficients $q_{ij}$ denote the weight of the interaction between the variables $x_i$ and $x_j$.

D-Wave Systems Inc. \cite{hu2019quantum} is a pioneering quantum computing company that utilizes QA to construct quantum computers. These state-of-the-art systems harness the principles of quantum mechanics to address complex problems, with a particular focus on optimization and sampling problems. D-Wave also provides the Ocean SDK, a Python-based open-source software development kit for the development of quantum applications. Access to the computational power of D-Wave’s quantum computer is provided through their Quantum Cloud Service. D-Wave offers a variety of solvers, including the D-Wave  QPU Solver, Leap Hybrid Solver, and several Classical solvers. These solvers take a problem formulated by the user and leverage the unique properties of a quantum system to find solutions.

\section{Proposed Approach}
\begin{figure}[!htb]
    \centering
    \tikz \node [scale=0.7, inner sep=5] {    
    \begin{tikzpicture}        
        \tikzset{
        block/.style= {draw, rectangle,rounded corners, align=center,minimum width=2cm,minimum height=1cm,text width= 2cm, fill=green!20},
        block1/.style= {draw, rectangle,rounded corners, align=center,minimum width=3 cm,minimum height=1cm,text width= 3cm, fill=green!20},
        line/.style = {draw, -latex'}
    }
\node[block1, node distance=1.5cm](raw_data){Raw Datasets};
\node[block1, below of=raw_data, node distance=1.5cm](data_cleaning){Data Cleaning};
\node[block1, below of=data_cleaning,   node distance=1.5cm](data_balancing){Data Balancing};
\node[block, below of=data_balancing, node distance=1.5cm,xshift=-3cm](training_set){Training Set};
\node[block, below of=data_balancing, node distance=1.5cm,xshift=3cm](testing_set){Testing Set};
\node[block1, below of=training_set, node distance=2cm](training_norm){Training Data Normalization};
\node[block1, below of=testing_set, node distance=2cm](testing_norm){Testing Data Normalization};

\node[block, below of=training_norm, node distance=2.5cm,xshift=3cm](qubo){QUBO Formulation};
\node[block, below of=qubo, node distance=1.5cm, xshift=-1.26cm](classical){Classical Solver};
\node[block, below of=qubo, node distance=1.5cm,xshift=1.26cm](quantum){QA Solver};

\node[above of = qubo, yshift= 0.1cm] (B) {Feature Subset selection};
\node[fit= (qubo) (classical) (quantum) (B), rounded corners, draw,inner sep=0.2cm, ] (FS_block){};

\node[block1, below of=classical, xshift= 1.2cm, node distance=1.5cm](Tr_f){Selected Features};

\node[block, below of=Tr_f, xshift= -1.5cm,node distance=2cm](Tr_f_data){ Training
Dataset with Selected Features};
\node[block, below of=Tr_f, xshift= 2cm, node distance=2cm](Ts_f_data){Testing
Dataset with Selected Features};

\node[block, below of=Tr_f_data, node distance=2.5cm,xshift=-1.5cm](ML){Machine Learning Algorithm};

\node[block, right of= ML, node distance=2cm, xshift=1.3cm](Model){Prediction Model};

\node[block, right of= Model, node distance=2cm,xshift=1.3cm](prediction){Prediction};

\path [line] (raw_data) -- (data_cleaning);
\path [line] (data_cleaning) -- (data_balancing);
\path [line] (data_balancing) -| (training_set);
\path [line] (data_balancing) -| (testing_set);
\path [line] (training_set.south) -| (training_norm);
\path [line] (testing_set.south) -| (testing_norm);
\path [line] (training_norm.east) -| (FS_block.north);

\path [line] (training_set.south) -| (training_norm);

\path [line] (qubo) -- (classical);
\path [line] (qubo) -- (quantum);
\path [line] (FS_block.south) -| (Tr_f);
\path [line] (training_norm.south) |- (Tr_f.west);
\path [line] (testing_norm.south) |- (Tr_f.east);
\path [line] (Tr_f_data) -| (ML);
\path [line] (Ts_f_data.west) -|(Model);
\path [line] (Tr_f) -- (Tr_f_data);
\path [line] (Tr_f) -- (Ts_f_data);
\path [line] (ML) -- (Model);
\path [line] (Model) -- (prediction);

\end{tikzpicture}
};
\caption{Workflow of proposed feature subset selection for software defect prediction }
\label{fig:block_diagram}
\end{figure}
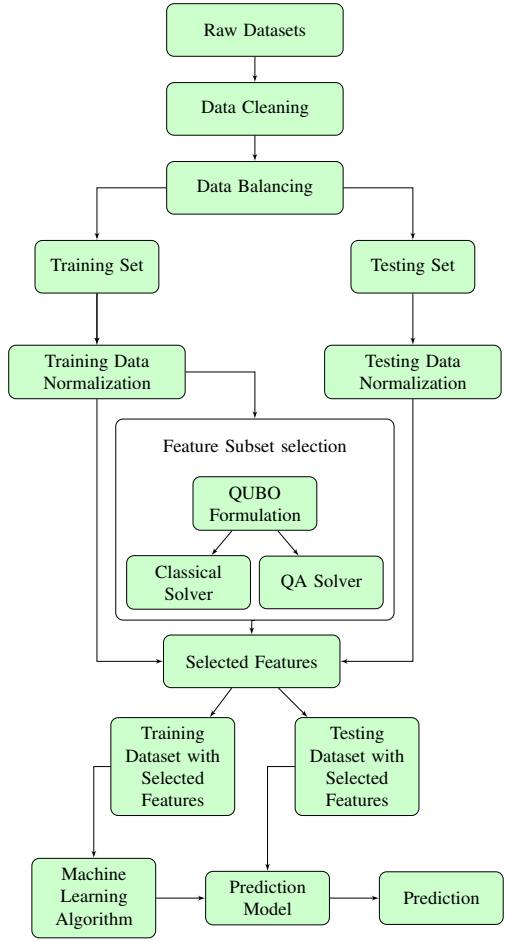

This section presents a detailed explanation of the process of employing an QA machine for feature subset selection in software defect prediction. The general workflow of the proposed approach is shown in Figure \ref{fig:block_diagram}. Initially, we gather publicly available software defect prediction datasets. As these datasets are raw, we undertake several preprocessing steps to clean the data. We tackle the class imbalance problem using an oversampling method. Following this, we partition the data into training and testing sets and apply data normalization to both sets. The feature subset selection is then performed on the training datasets. We employ filter-based QUBO functions to identify the appropriate feature subset. We utilize both a  D-Wave classical solver and the  QA solver. The training dataset, with the selected features, is used to construct the model using a Support Vector Machine (SVM) classifier. The testing dataset, with the selected features, is used to evaluate the prediction performance. Finally, we present the performance results of both the classical and QA solvers in addressing feature subset selection for the software defect prediction task. 

\textbf{Dataset description:} For the software defect prediction task, we utilized 24 publicly available datasets sourced from three different project repositories, including AEEEM \cite{D’Ambros2012}, JIRA \cite{Yatish8811982}, and NASA \cite{6464273Shepperd}. Table \ref{tab:datasets} shows the software defect prediction datasets, its project repository containing, the number of features of the dataset, the number of samples and the class distribution. The class distribution of each data set shows the number of instances that are buggy (1) and nonbuggy (0), with nonbuggy presenting much more than buggy (imbalance datasets). The features of the datasets usually include different software metrics, including code metrics and process metrics. Code metrics such as lines of code and cyclomatic complexity are extracted directly from the source code. On the other hand, process metrics are derived from historical data stored in various software repositories, including version control and issue tracking systems. 
For convenience, the original dataset names of JIRA projects such as wicket-1.3.0-beta2, activemq-5.0.0, groovy-1\_6\_BETA\_1, hbase-0.94.0, and hive-0.9.0 have been shortened to wicket, activemq, groovy, hbase, and hive, respectively.

\begin{table}[htb!]
\caption{Software Defect prediction datasets}
\centering
\label{tab:datasets}
\begin{tabular}{|p{0.9cm}|p{1cm}|p{1cm}|p{1cm}|p{2cm}|}
\hline
\textbf{Projects  }                          & \textbf{Dataset }                 & \textbf{Number of  Features} & \textbf{Number of  Samples} & \textbf{Class   Distribution }         \\ \hline
\multirow{5}{*}{AEEEM}              & EQ                   & 61           & 324         & {0: 195, 1: 129}  \\ \cline{2-5} 
                                    & JDT                  & 61           & 997         & {0: 791, 1: 206}  \\ \cline{2-5} 
                                    & Lucene               & 61           & 691         & {0: 627, 1: 64}   \\ \cline{2-5} 
                                    & Mylyn                & 61           & 1862        & {0: 1617, 1: 245} \\ \cline{2-5} 
                                    & PDE                  & 61           & 1497        & {0: 1288, 1: 209} \\ \hline
\multirow{6}{*}{JIRA}               & activemq& 65           & 1884        & {0: 1591, 1: 17293}           \\ \cline{2-5} 
                                    & groovy    & 65           & 821         & {0: 751, 1: 70}               \\ \cline{2-5} 
                                    & hbase        & 65           & 1059        & {0: 841, 1: 218}              \\ \cline{2-5} 
                                    & hive         & 65           & 1416        & {0: 1133, 1: 283}             \\ \cline{2-5} 
                                    & jruby         & 65           & 731         & {0: 644, 1: 87}               \\ \cline{2-5} 
                                    & wicket   & 65           & 1763        & {0 : 1633, 1: 130}             \\ \hline
\multirow{13}{*}{NASA}              & CM1                  & 40           & 505         & {0: 457, 1: 48}           \\ \cline{2-5} 
                                    & JM1                  & 21           & 10878       & {0: 8776, 1: 2102}        \\ \cline{2-5} 
                                    & KC1                  & 21           & 2107        & {0: 1782, 1: 325}         \\ \cline{2-5} 
                                    & KC3                  & 40           & 458         & {0: 415, 1: 43}           \\ \cline{2-5} 
                                    & KC4                  & 40           & 125         & {0: 64, 1: 61}            \\ \cline{2-5} 
                                    & MC1                  & 39           & 9466        & {0: 9398, 1: 68}          \\ \cline{2-5} 
                                    & MC2                  & 40           & 161         & {0: 109, 1: 52}           \\ \cline{2-5} 
                                    & MW1                  & 40           & 403         & {0: 372, 1: 31}           \\ \cline{2-5} 
                                    & PC1                  & 40           & 1107        & {0: 1031, 1: 76}          \\ \cline{2-5} 
                                    & PC2                  & 40           & 5589        & {0: 5566, 1: 23}          \\ \cline{2-5} 
                                    & PC3                  & 40           & 1563        & {0: 1403, 1: 160}         \\ \cline{2-5} 
                                    & PC4                  & 40           & 1458        & {0: 1280, 1: 178}         \\ \cline{2-5} 
                                    & PC5                  & 39           & 17186       & {0: 16670, 1: 516}        \\ \hline
\end{tabular}
\end{table}


\textbf{Data Preprocessing:} The collected raw datasets often contain missing values, duplicate samples, or features with constant values. We handled missing values by imputing them with the mean value. We eliminated duplicate entries and features with constant values to prevent bias in model training. As observed in the dataset description, the distribution of classes in the code projects is imbalanced. There are fewer instances of buggy code than the correct ones. This imbalance can lead classification algorithms to ignore the smaller class (defect-prone code), resulting in overfitting. To address this imbalance, we used oversampling methods based on the synthetic minority oversampling technique (SMOTE) to prepare the dataset. We partitioned the dataset into training and testing sets, followed by Min-Max normalization to scale the data within the [0, 1] range for both sets.

To address feature subset selection in software defect prediction using the QA solver, this problem needs to be transformed into QUBO form. In feature subset selection, each feature of a feature subset can be represented as a binary decision, where 0 indicates the feature is not selected and 1 indicates the feature is selected. This representation of input is equivalent to QUBO input vectors, which are binary variables. In order to transform feature subset selection into QUBO, we need to define the objective function equivalent to QUBO formulation. To perform this, we use a MI-based filter feature subset selection approach and define a QUBO objective function. We have answered our RQ1 here.

We have to find the best subset of features $X_{best}$ that minimizes the objective function of a feature vector $X=\{x_1,x_2,...,x_n\}$. In other words, finding the best subset that returns  the minimum value for the objective function. We can consider two terms: the linear term encompasses all interactions between individual feature terms with the target variable, and the quadratic term is the sum of interactions between all feature pairs. The intuition is that a feature subset is better if its feature-target interaction is stronger than its feature-feature interaction.  That is, the best feature subset minimizes the (\text{Quadratic Term}-\text{Linear Term}) value.
We incorporate two coefficients, denoted as $\alpha$ and $\beta$  to regulate the linear and quadratic terms, respectively, in order to manage both terms effectively. The QUBO equation can be represented as follow.

\begin{equation}
    \text{Minimize} \quad f(X) =- \alpha \sum_{i} MI_{ii}x_i +\beta \sum_{i<j} MI_{ij}x_ix_j
\end{equation}

Where, in linear terms, $MI_{ii}$ represents the MI between the target and a feature, while $MI_{ij}$ denotes the MI between two features. The parameters $\alpha$ and $\beta$ serve as a relative weight, such that $\alpha+\beta=1$.

The MI-based objective function now represents a QUBO problem, which can be addressed using both QA systems and classical machines. For the QA machine, the problem is further converted into qubits of the QA through a process known as minor embedding \cite{choi2011minor}. This method maps the logical qubits of the QUBO problem onto the physical qubits of the QA machine. These solvers tackle optimization problems by identifying the ground state of a given Hamiltonian function \cite{rizk2019deep}, which corresponds to the final solution set of features for predicting software defects.

In this feature subset selection task, we use Leap's QPU solver, which can be accessed via D-Wave’s Leap quantum cloud service. We submit the problem through this API, execute the problem, and collect the results. It is worth noting that the software can automatically handle minor-embedding for the QUBO objective functions. In addition, as a baseline approach, we use a classical solver that runs on our local CPU. 

\textbf{Experimental Design:} In the process of data partitioning, we employ a stratified random sampling approach to create a training set and a testing set, comprising 80\% and 20\% of the data, respectively. The model of choice for both training and testing is an SVM with a linear kernel function. To ensure the consistency and reproducibility of our results, a fixed seed is used for the random number generator. We utilized the Ocean SDK provided by D-Wave Systems for QA-based feature subset selection. To implement the classical solver, we use the '\textit{dwave-samplers}' package that contains different classical solvers. In this experiment, we use SA as it is one of the popular and efficient classical solvers. The solver parameters are left with their default settings, with the exception of the number of repetitions, which is explicitly set to $num\_reads=100$. For QA solver, we utilize D-Wave Ocean’s '\textit{dwave-system}' API, which allows users to access Leap’s cloud-based hybrid solvers and the QPU solver smoothly.   For our classical computing needs, we utilize a robust Linux server. This server is powered by an Intel® Xeon® Platinum 8356H CPU with 64 cores, operating at a speed of 3.90 GHz and featuring a CPU frequency of approximately 4400 MHz. In Quantum computing, D-Wave  Leap automatically selects the available QPU at that time of computing. In our MI-based QUBO objective, it is crucial to adjust the parameters to achieve a better optimal solution. We fine-tune these parameters using a trial-and-error approach, with $\alpha$ set to 0.98 and $\beta$ set to 0.02, respectively.

Along with feature subset selection, we also consider the scenario without feature subset selection to determine the software defect prediction. This is to check whether feature subset selection can reduce the number of features while still providing competitive accuracy. In this experiment, we consider performance metrics such as feature subset size and execution time, along with the accuracy and F-score performance of the prediction model.

\section{Result and Discussion}
Table \ref{tab:feature_selection} presents the performance of D-Wave QPU and classical solvers in reducing feature cardinality for software defect prediction.  Across nearly all datasets, the two solvers demonstrate significant feature reduction. This reduction is most pronounced for datasets with large initial feature sets, such as AEEEM and the JIRA project. The solvers exhibit comparable feature subset selection performance.  These results support an affirmative answer to RQ2: the QA solver can significantly reduce feature cardinality in most cases.

\begin{table}[ht]
\centering
\caption{Comparison of Feature Subset Selection Results Across Datasets: Original vs. Selected Features Using D-Wave Classical and QPU Solvers}
\label{tab:feature_selection}
\begin{tabular}{|c|c|c|c|c|}
\hline
\textbf{Projects }              & \textbf{Dataset  }            & \begin{tabular}[c]{@{}l@{}}\textbf{Original} \\ \textbf{Features}\end{tabular}  & \textbf{Classical Solver}  & \textbf{QPU Solver} \\ \hline

\hline
\multirow{5}{*}{AEEEM} & EQ       & 61 & 37 &  39 \\ \cline{2-5} 
                       & JDT      & 61 & 43 &  43 \\ \cline{2-5} 
                       & Lucene   & 61 & 40 &  40 \\ \cline{2-5} 
                       & Mylyn    & 61 & 36 &  38 \\ \cline{2-5} 
                       & PDE      & 61 & 39 & 39 \\ \hline
\multirow{6}{*}{JIRA}  & activemq & 65 & 43 & 44 \\ \cline{2-5} 
                       & groovy   & 65 & 43 & 44 \\ \cline{2-5} 
                       & hbase    & 65 & 42 &  43 \\ \cline{2-5} 
                       & hive     & 65 & 43 &  42 \\ \cline{2-5} 
                       & jruby    & 65 & 44 &  46 \\ \cline{2-5} 
                       & wicket   & 65 & 43 & 45 \\ \hline
\multirow{13}{*}{NASA} & CM1      & 40 & 34 & 35 \\ \cline{2-5} 
                       & JM1      & 21 & 21 &  21 \\ \cline{2-5} 
                       & KC1      & 21 & 20 &  20 \\ \cline{2-5} 
                       & KC3      & 40 & 37 &  36 \\ \cline{2-5} 
                       & KC4      & 40 & 11 & 13 \\ \cline{2-5} 
                       & MC1      & 39 & 32 & 33 \\ \cline{2-5} 
                       & MC2      & 40 & 32 & 33 \\ \cline{2-5} 
                       & MW1      & 40 & 33 &  32 \\ \cline{2-5} 
                       & PC1      & 40 & 34 &  34 \\ \cline{2-5} 
                       & PC2      & 40 & 34 &  34 \\ \cline{2-5} 
                       & PC3      & 40 & 35 &  35 \\ \cline{2-5} 
                       & PC4      & 40 & 33 &  33 \\ \cline{2-5} 
                       & PC5      & 39 & 35 &  35 \\ \hline
\end{tabular}
\end{table}

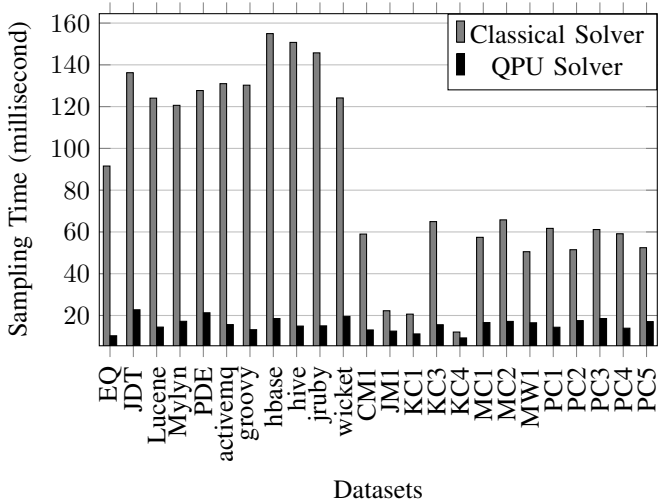
\begin{figure}[ht]
	\centering
	\pgfplotsset{compat=1.11,
		/pgfplots/ybar legend/.style={
			/pgfplots/legend image code/.code={%
				\draw[##1,/tikz/.cd,yshift=-0.25em]
				(0cm,0cm) rectangle (3pt,0.8em);},
		},
	}
	\pgfplotstableread[col sep=comma,]{sampling-time.csv}\datatable
	\begin{tikzpicture}
	\begin{axis}[width=9cm,height=6cm,  
	enlargelimits=0.03,
	ybar=0.1*\pgflinewidth,
	enlarge x limits=0.02,
	ymin=10.0, 
	ymax=160.0,
	ytick={0,20,40,60,80,100,120,140,160},
	bar width=2.5pt,
	xtick=data,
	x tick label style={rotate=90,anchor=east},
	xticklabels from table={\datatable}{Datasets},
	axis background/.style={fill=white},
	ymajorgrids = true,
	major x tick style = transparent,
	legend style={legend columns=1, at={(1.0, 1.0)}},
	xlabel={ Datasets},
	ylabel={Sampling Time (millisecond)},
	]
	\addplot [fill=gray]  table [x expr=\coordindex, y=Classical Solver
	]{\datatable};
	\addplot[fill=black]  table [x expr=\coordindex, y=QPU Solver]{\datatable};
	\legend{Classical Solver
		,QPU Solver}
	\end{axis}
	\end{tikzpicture}
	\caption{Comparison of D-wave classical and QPU solver sampling times for feature subset selection in software defect prediction}
	\label{fig:time}
\end{figure}

To address RQ3, we performed a computational time analysis to evaluate the performance of the D-Wave QPU solver compared to a classical solver for selection of feature subsets in software defect prediction. The results are shown in Figure \ref{fig:time}. We specifically focused on sampling time, measured in milliseconds, which reflects the duration each system requires to converge toward an optimal solution. The results demonstrate a consistent speed advantage for the QPU solver in all datasets.  In particular, for the AEEM and JIRA Project datasets, the QPU solver found optimal feature subsets significantly faster than the classical solver – on average, less than 25 milliseconds for the QPU compared to over 90 milliseconds for the classical solver. Based on our findings, we conclude that QA demonstrates a superior speed compared to the classical solver in the task of selecting feature subsets for software defects prediction.

Finally, to answer the research question RQ4, we evaluated the performance of software defect prediction models with and without feature subset selection. Table \ref{tab:accuracy-results} shows the accuracy and F1 scores calculated for each dataset, with the best scores highlighted in bold. Similar accuracy and F1 scores suggest a relatively balanced distribution of defective and non-defective software modules. The QPU solver exhibited slightly better accuracy and F1 scores compared to the classical solver for most of the datasets. Moreover, models utilizing feature subset selection demonstrated competitive classification accuracy and F1 scores compared to models without feature subset selection.

\begin{table}[ht]
\caption{Comparison of Accuracy and F1 Scores for Software Defect Prediction Models With and Without Feature Subset Selection Using D-Wave's Classical and QPU Solvers}
\centering
\label{tab:accuracy-results}
\begin{tabular}{|l|cc|cc|cc|}
\hline
\multirow{2}{*}{\textbf{Projects}} & \multicolumn{2}{c|}{\textbf{All features}}                            & \multicolumn{2}{c|}{\textbf{Classical Solver}}                            & \multicolumn{2}{c|}{\textbf{QPU Solver}}                             \\ \cline{2-7} 
                                   & \multicolumn{1}{l|}{ACC}       & \multicolumn{1}{l|}{F1}       & \multicolumn{1}{l|}{ACC}       & \multicolumn{1}{l|}{F1}       & \multicolumn{1}{l|}{ACC}       & \multicolumn{1}{l|}{F1}       \\ \hline
\multirow{5}{*}{AEEEM}             & \multicolumn{1}{c|}{\textbf{78.21}} & \textbf{78.21}                & \multicolumn{1}{c|}{78.00}          & 76.92                         & \multicolumn{1}{c|}{\textbf{78.21}} & 78.11                         \\ \cline{2-7} 
                                   & \multicolumn{1}{c|}{\textbf{82.33}} & \textbf{82.30}                & \multicolumn{1}{c|}{81.39}          & 81.35                         & \multicolumn{1}{c|}{81.70}          & 81.66                         \\ \cline{2-7} 
                                   & \multicolumn{1}{c|}{84.46}          & 84.46                         & \multicolumn{1}{c|}{82.07}          & 82.07                         & \multicolumn{1}{c|}{\textbf{85.66}} & \textbf{85.62}                \\ \cline{2-7} 
                                   & \multicolumn{1}{c|}{\textbf{82.84}} & \textbf{82.78}                & \multicolumn{1}{c|}{80.68}          & 80.65                         & \multicolumn{1}{c|}{80.83}          & 80.81                         \\ \cline{2-7} 
                                   & \multicolumn{1}{c|}{79.07}          & 79.04                         & \multicolumn{1}{c|}{\textbf{79.26}} & \textbf{79.23}                & \multicolumn{1}{c|}{77.71}          & 77.69                         \\ \hline
\multirow{6}{*}{JIRA}              & \multicolumn{1}{c|}{\textbf{89.48}} & \textbf{89.48}                & \multicolumn{1}{c|}{86.03}          & 85.97                         & \multicolumn{1}{c|}{86.19}          & 86.14                         \\ \cline{2-7} 
                                   & \multicolumn{1}{c|}{83.72}          & 83.70                         & \multicolumn{1}{c|}{\textbf{84.05}} & \textbf{84.03}                & \multicolumn{1}{c|}{\textbf{84.05}} & \textbf{84.03}                \\ \cline{2-7} 
                                   & \multicolumn{1}{c|}{\textbf{79.82}} & \textbf{79.70}                & \multicolumn{1}{c|}{78.04}          & 77.83                         & \multicolumn{1}{c|}{76.85}          & 76.68                         \\ \cline{2-7} 
                                   & \multicolumn{1}{c|}{\textbf{80.18}} & \textbf{80.06}                & \multicolumn{1}{c|}{74.45}          & 74.29                         & \multicolumn{1}{c|}{73.79}          & 73.53                         \\ \cline{2-7} 
                                   & \multicolumn{1}{c|}{\textbf{87.21}} & \textbf{87.21}                & \multicolumn{1}{c|}{84.11}          & 84.10                         & \multicolumn{1}{c|}{86.05}          & 86.05                         \\ \cline{2-7} 
                                   & \multicolumn{1}{c|}{87.00}          & 87.00                         & \multicolumn{1}{c|}{\textbf{87.46}} & \textbf{87.46}                & \multicolumn{1}{c|}{87.00}          & 87.00                         \\ \hline
\multirow{13}{*}{NASA}             & \multicolumn{1}{c|}{\textbf{81.97}} & \textbf{82.05}                & \multicolumn{1}{c|}{81.97}          & 81.99                         & \multicolumn{1}{c|}{\textbf{81.97}} & \textbf{82.05}                \\ \cline{2-7} 
                                   & \multicolumn{1}{c|}{\textbf{65.25}} & \textbf{64.84}                & \multicolumn{1}{c|}{\textbf{65.25}} & \textbf{64.84}                & \multicolumn{1}{c|}{\textbf{65.25}} & \textbf{64.84}                \\ \cline{2-7} 
                                   & \multicolumn{1}{c|}{\textbf{71.95}} & \textbf{71.94}                & \multicolumn{1}{c|}{71.11}          & 71.11                         & \multicolumn{1}{c|}{71.11}          & 71.11                         \\ \cline{2-7} 
                                   & \multicolumn{1}{c|}{75.90}          & 75.59                         & \multicolumn{1}{c|}{\textbf{77.71}} & \textbf{77.53}                & \multicolumn{1}{c|}{75.90}          & 75.67                         \\ \cline{2-7} 
                                   & \multicolumn{1}{c|}{\textbf{76.92}} & \textbf{76.92}                & \multicolumn{1}{c|}{\textbf{76.92}} & \textbf{76.92}                & \multicolumn{1}{c|}{\textbf{76.92}} & \textbf{76.92}                \\ \cline{2-7} 
                                   & \multicolumn{1}{c|}{\textbf{94.95}} & \textbf{94.94}                & \multicolumn{1}{c|}{94.60}          & 94.59                         & \multicolumn{1}{c|}{94.87}          & 94.86                         \\ \cline{2-7} 
                                   & \multicolumn{1}{c|}{84.09}          & 84.17                         & \multicolumn{1}{c|}{\textbf{86.36}} & \textbf{86.48}                & \multicolumn{1}{c|}{\textbf{86.36}} & \textbf{86.48}                \\ \cline{2-7} 
                                   & \multicolumn{1}{c|}{\textbf{80.54}} & \textbf{80.70}                & \multicolumn{1}{c|}{79.87}          & 80.01                         & \multicolumn{1}{c|}{79.87}          & 80.01                         \\ \cline{2-7} 
                                   & \multicolumn{1}{c|}{81.36}          & 81.33                         & \multicolumn{1}{c|}{\textbf{82.32}} & \textbf{82.33}                & \multicolumn{1}{c|}{\textbf{82.32}} & \textbf{82.33}                \\ \cline{2-7} 
                                   & \multicolumn{1}{c|}{\textbf{91.92}} & \textbf{91.92}                & \multicolumn{1}{c|}{91.69}          & 91.69                         & \multicolumn{1}{c|}{91.69}          & 91.69                         \\ \cline{2-7} 
                                   & \multicolumn{1}{c|}{\textbf{84.34}} & \textbf{84.35}                & \multicolumn{1}{c|}{83.10}          & 83.10                         & \multicolumn{1}{c|}{83.10}          & 83.10                         \\ \cline{2-7} 
                                   & \multicolumn{1}{c|}{90.04}          & 89.91                         & \multicolumn{1}{c|}{\textbf{90.23}} & \textbf{90.10}                & \multicolumn{1}{c|}{\textbf{90.23}} & \textbf{90.10}                \\ \cline{2-7} 
                                   & \multicolumn{1}{c|}{93.78}          & 93.78                         & \multicolumn{1}{c|}{\textbf{93.84}} & \textbf{93.84}                & \multicolumn{1}{c|}{\textbf{93.84}} & \textbf{93.84}                \\ \hline
\end{tabular}
\end{table}

\section{Threats to Validity}



Our study employed three open-source projects containing 24 datasets. To further generalize our findings, future investigations should include a broader range of software defect datasets, particularly those with a larger number of features.  We used SVM classifiers for prediction performance in this study, but we plan to explore other classifiers in the future to understand their impact on software defect prediction outcomes. Due to limited access to the D-Wave QA machine, the experimental results presented here are based on a single run. Replicating experiments multiple times would enhance the generalizability of our findings. While several classical solvers are available, we chose a SA based solver due to its popularity and frequent use. Our study employed MI-based feature subset selection as the QUBO objective function, as it can capture both linear and non-linear relationships between variables.  To mitigate the threat of parameter sensitivity, we configured the QUBO parameters based on preliminary experiments. Since the QUBO objective function plays a crucial role in finding the appropriate feature subset, considering other objectives could impact the results.

\section{Related Works}
Feature subset selection has been extensively studied for classification tasks  and software defect prediction \cite{dokeroglu2022comprehensive}, with most approaches relying on classical computers for identifying relevant feature subsets. Recently, investigations have begun to explore how quantum computing could be utilized for feature subset selection \cite{milne2018optimal}. These investigations often involve simple toy datasets or simulations on classical computers.  However, research on QA machines or quantum computers specifically for software defect prediction still needs to be explored.

In \cite{tanahashi2018global}, feature selection is mapped into the Mutual Information-Based Feature Selection (MIFS) QUBO objective and solved using a D-Wave solver and classical solver. Notably, the quantum solver outperforms the classical counterpart. In the paper \cite{otgonbaatar2021quantum}, the authors  address the task of identifying an appropriate feature subset for hyper-spectral image classification. They utilize D-Wave QA machines, outperforming their classical counterparts in identifying highly informative bands for characterizing specific classes.  

Quantum computing has demonstrated potential in outperforming classical methods for feature subset selection in various domains; however, its application to software defect prediction is still largely unexplored. This study aims to address this gap, potentially paving the way for more efficient and accurate defect prediction models.


\section{Conclusion}
In this study, we explored the performance of a QA-based feature subset selection technique for software defect prediction.  We employed the D-Wave QPU solver as QA solver for feature subset selection, which we accessed via the D-Wave Leap cloud service.  For comparison, we also used a D-Wave classical solver running on the traditional CPU. We utilized a MI-based feature subset selection as a QUBO objective for feature subset selection. Our experiments, conducted on 24 datasets from three different projects, showed that the QA solver is as effective as the classical solver in reducing the number of features. However, the QA solver requires less computation time than the classical solver to find the optimal feature subset.  In most cases, the QA solver based feature subset selection combined with a classification model improved the software defect prediction performance more than its classical counterpart with the classification model. In future work, we plan to evaluate the efficacy of gate-model quantum computing for the same task. In addition, we intend to use other feature subset selection techniques and larger-scale software defect prediction datasets to further investigate the performance of quantum computing. It is important to note that quantum computing is still in its early stages, and its benefits are likely to become more evident in the near future. 

\section*{Acknowledgment}
This research is supported in part by the Natural Sciences and Engineering Research Council of Canada (NSERC) Discovery Grants program, and by the industry-stream NSERC CREATE in Software Analytics Research (SOAR). 

\bibliographystyle{plainnat} 
\setcitestyle{numbers}
\footnotesize
\bibliography{sample}

\end{document}